\documentclass[12pt,a4paper]{iopart}
\usepackage{multibib}
\usepackage{xcolor}
\usepackage{amssymb}
\usepackage{graphicx}
\usepackage{multirow}
\usepackage{subfig,cite}
\usepackage[margin=1.1 in]{geometry}

\begin{document}
\title[Bootstrapping the Kronig-Penney Model]{Bootstrapping the Kronig-Penney Model}

\author{Matthew J. Blacker$^{(a,c)}$, Arpan Bhattacharyya$^b$  and Aritra Banerjee$^c$}

\address{$^a$ Department of Applied Mathematics and Theoretical Physics,
University of Cambridge, Cambridge CB3 0WA, United Kingdom\\

\medskip
$^b$ Indian Institute of Technology, Gandhinagar, Gujarat-382355, India\\
\medskip
$^c$ Okinawa Institute of Science and Technology,
1919-1 Tancha, Onna-son, Okinawa 904-0495, Japan
}
\ead{mjb318@cam.ac.uk, abhattacharyya@iitgn.ac.in, aritra.banerjee@oist.jp}

%\author{Arpan Bhattacharyya}

%\address{Indian Institute of Technology, Gandhinagar, Gujarat-382355, India}
%\ead{abhattacharyya@iitgn.ac.in}

%\author{Aritra Banerjee}

%\address{Okinawa Institute of Science and Technology,
%1919-1 Tancha, Onna-son, Okinawa 904-0495, Japan}
%\ead{aritra.banerjee@oist.jp}

\vspace{10pt}
\begin{indented}
\item[]
%\today
\end{indented}

\begin{abstract}
Recently, bootstrap methods from conformal field theory have been adapted for studying the energy spectrum of various quantum mechanical systems. In this paper, we consider the application of these methods in obtaining the spectrum from the Schrödinger equation with periodic potentials, paying particular attention to the Kronig-Penney model of a particle in a one-dimensional lattice. With an appropriate choice of operator basis involving position and momenta, we find that the bootstrap approach efficiently computes the band gaps of the energy spectrum but has trouble effectively constraining the minimum energy. We show how applying more complex constraints involving higher powers of momenta can potentially remedy such a problem. We also propose an approach for analytically constructing the dispersion relation associated with the Bloch momentum of the system.
\end{abstract}

\section{Introduction}

Most quantum mechanical systems do not possess an analytic solution, even in one dimension. Merely brute force numerical methods for such problems can be costly; hence, symmetry principles that enhance the efficiency of such a process have historically proved useful. In this regard, inspired by \cite{lin_bootstraps_2020}, the bootstrapping methodology of conformal field theory\footnote{To be familiarised with state of the art for CFT bootstrap techniques, the reader is directed for example to \cite{poland_conformal_2019,bissi_selected_2022} and the references within. This list is in no way exhaustive.} was recently adapted in \cite{han_bootstrapping_2020} in a bid to numerically solve such classes of systems that are not analytically soluble or are hard to do so even numerically.
\medskip

The central philosophy of their bootstrapping approach is as follows: given a set of initial data for a Hamiltonian quantum system (such as energy eigenvalues), other statistical moments of quantum operators can be computed using symmetries of the system. These moments of different order are, in turn, related to each other via recursion relations, effectively reducing the search space for independent data. To determine whether the initial data is physically viable (i.e. whether it is consistent with a square-integrable eigenstate of the whole Schr\"{o}dinger equation), one then enforces an adequate set of constraints upon the computed statistical moments, such as requiring the positivity of quadratic operators. By computing more and more statistical moments and testing them against the given constraints, the goal is that the set of allowed initial data points will converge to the system's actual solution.
\medskip 

The bootstrapping methodology of \cite{han_bootstrapping_2020}, implemented in that work as a tool to solve matrix models and matrix quantum mechanics, has recently been applied to numerically solve various quantum mechanical problems \cite{aikawa_bootstrap_2022,aikawa_application_2022,berenstein_bootstrapping_2021,berenstein_bootstrapping_2022,berenstein_anomalous_2022,bhattacharya_numerical_2021,hu_different_2022,khan_bootstrapping_2022,morita_bootstraping_2022,tchoumakov_bootstrapping_2021,nakayama_bootstrapping_2022,du_bootstrapping_2022,bai_bootstrapping_2022,li_null_2022}. In particular, the procedure has been used to find the energy spectrum of physical systems, and has proved most effective at finding lower energy modes. Furthermore, it has been observed that the convergence speed of the bootstrap is improved by computing statistical moments that have the same symmetries as the total system. That is, the efficiency of the bootstrap increases by, for example, considering creation and annihilation operators for a diagonalisable system \cite{aikawa_bootstrap_2022}, or periodic moments for a periodic potential \cite{aikawa_application_2022,berenstein_bootstrapping_2022,tchoumakov_bootstrapping_2021}.
\medskip

With such a potentially powerful algorithm by our side, we must test out the effectiveness of the system by employing it to solve already tractable sectors. Ideally, one hopes to use the bootstrap as a trained black-box to readily spew out the spectrum of a quantum system with certain symmetries, and the training requires it to go through many familiar problems. Our present work is just another small step towards that goal.

\medskip
More specifically, as we have already mentioned, work has recently begun in applying the bootstrapping procedure to quantum systems with periodic potentials, which have several applications, such as in solid state physics \cite{harper_single_1955,aubry_analyticity_1980,marsal_topological_2020}. In particular, \cite{aikawa_application_2022,berenstein_bootstrapping_2022,tchoumakov_bootstrapping_2021} have considered a cosine potential (the Mathieu problem), and successfully reproduced the expected energy spectrum. However, reconstructing the dispersion relation for the quasimomentum has proved challenging. In \cite{tchoumakov_bootstrapping_2021}, the authors had success using a statistical approach, approximating the dispersion relation from a probability distribution for momentum reconstructed from a finite number of moments.
\medskip

 In this paper, we continue the investigation of the quantum mechanical bootstrap as applied to periodic potentials. Starting with a generic periodic potential, we concentrate on the Kronig-Penney model \cite{kronig_quantum_1931}, a simplified model of a particle in a one-dimensional lattice where a series of periodic delta spikes model the crystal structure (the Dirac comb), which has applications from models of graphene \cite{masir_magnetic_2009} to ultra-cold atoms \cite{negretti_generalized_2014}, to name a few. We choose a general spatially periodic operator basis to compute our recursion relation and reconstruct the probability density function using that relation\footnote{Note this analytical trick of using moment recursion relations was already discussed many decades ago in \cite{banerjee_transition_1977}.}. A numerical search for allowed energy eigenvalues restricted by the positivity constraint for the generic operator basis gives rise to the discrete band structure associated with the problem. We find the convergence of bootstrapped energy bands to the analytic ones depends explicitly on the structure of the operator basis chosen. We also sketch a way to extract the exact dispersion relation for the band structure using our bootstrap data.
\medskip

The rest of the paper is organised as follows: In Section \ref{SecBootIdentities}, we review the general bootstrapping identities and use them to construct a recursion relation first for a general periodic potential and then specialise in the Kronig-Penney model. In Section \ref{SecNumericImplementation} we review the positivity constraint for periodic operators, numerically reproduce the band gaps of the known solution to the Kronig-Penney model, and discuss how to obtain the full energy spectrum from the positivity constraint for a periodic potential. In Section \ref{SecDispersionRelation}, we find that an analytic approach can reconstruct the exact dispersion relation of the model. We conclude with a teaser of the road ahead. The appendices contain some extra details and neat examples of anomalies appearing in recursion relations for completeness. 

\section{Bootstrapping identities for a periodic potential}
\label{SecBootIdentities}
We begin with a summary of how to take stock of the initial data associated with a bootstrap problem, mostly following the treatise of \cite{han_bootstrapping_2020}.
We begin with the general one-dimensional Hamiltonian with a space-dependent potential
\begin{eqnarray}
H = p^2 + V(x),
\end{eqnarray}
defined over the interval $\lbrace x_1 \leq x \leq x_2\rbrace$, where $\left[ \hat{x}, \hat{p} \right] = i$, and we use throughout the convention $\hbar = 1$ and $2m = 1$. Consider an eigenstate $|\psi \rangle$ of the Hamiltonian $\hat{H}$ with energy $E$, defined in the position basis $|x\rangle$ as
\begin{eqnarray}
| \psi \rangle = \int_{x_1}^{x_2} dx \psi(x) |x \rangle.
\end{eqnarray}
In the eigenstate $| \psi \rangle$, the expectation values of position-space functions $f(x)$ are defined as
\begin{eqnarray}
\left \langle f(x) \right \rangle = \int_{x_1}^{x_2} dx f(x) \rho(x),
\end{eqnarray}
where $\rho(x) = \left| \psi(x) \right|^2$ is the wavefunction density. The first step of the bootstrapping procedure is to construct a relation for moments of the distribution $\rho(x)$ using two identities. The first identity is that, for an operator $\hat{O}$, the following average vanishes, albeit up to an anomaly:
\begin{eqnarray}
\left \langle \left[ \hat{H}, \hat{O} \right] \right \rangle + \mathcal{A}_{\hat{O}} = 0, \label{EqConstraint1}
\end{eqnarray}
where $\mathcal{A}_{\hat{O}}$ is the anomaly defined in \cite{berenstein_anomalous_2022} as 
\begin{eqnarray}
\mathcal{A}_O = \left \langle \left( H^{\dagger} - H \right) O \right \rangle_{\psi} = \int_{x_1}^{x_2} dx \left[ \left( \hat{H} \psi(x) \right)^{\dagger} - \psi^{\dagger} (x) \hat{H} \right] \hat{O} \psi(x). \label{EqAnomalyDefinition}
\end{eqnarray}
When expectation values are computed over a finite domain, the anomaly generates contact terms in the recursion relation \cite{berenstein_anomalous_2022}. Mathematically, it corresponds to the difference between computing the expectation values by approaching the boundaries of the domain from the left or the right.
\medskip 

Using equation (\ref{EqConstraint1}), and choosing two classes of operators, namely $\hat{O} = f(x)$ and $\hat{O} = f(x)p$, we obtain the following relations:
\numparts
\begin{eqnarray}
 - \langle f'' \rangle - 2i \langle f' p \rangle + \mathcal{A}_{f} = 0, \label{EqFCon1}\\
    - \langle f'' p \rangle - 2i \langle f' p^2 \rangle + i \langle f V' \rangle + \mathcal{A}_{fp} = 0, \label{EqFPCon1}
\end{eqnarray}
\endnumparts
having used the identity $\left[ p, f \right] = -i f'$ and the notation $f^{(n)} = \partial^n f/ \partial x^n$. The second identity we need is that as $|\psi \rangle$ is an energy eigenstate of the Hamiltonian, 
\begin{eqnarray}
    \left \langle \hat{O} \hat{H}\right \rangle = E \left \langle \hat{O} \right \rangle, \label{EqConstraint2}
\end{eqnarray}
from which (for $\hat{O} = f$) we obtain the relation
\begin{eqnarray}
    \langle fp^2 \rangle + \langle f V \rangle - E \langle f \rangle = 0. \label{EqFCon2} 
\end{eqnarray}
By relating $f$ to its derivatives, from equations (\ref{EqFCon1}), (\ref{EqFPCon1}) and (\ref{EqFCon2}), we can construct a relation between moments of the probability distribution.

\subsection{A general periodic potential}
For a periodic potential of period $a$, the domain of interest becomes $\lbrace - a/2 \leq x \leq a/2 \rbrace$ and the wavefunction satisfies the Dirichlet boundary condition $\psi(-a/2) = \psi(a/2)$. A stronger constraint is actually imposed by the Bloch theorem, which states that the wavefunctions of energy eigenstates of a periodic lattice can be written as $\psi(x) = \exp \left( i k x \right) u(x)$. Here, $k$ is a wavevector bounded by $\lbrace -\pi/a \leq k \leq \pi/a \rbrace$, and $u(x) = u(x+a)$ i.e. a function with the same periodicity of the lattice.
\medskip

For our choice of $f$, it will prove convenient to consider a spatially periodic function. In particular, we pick $f = t_n = \exp \left( i 2\pi n x/ a \right)$, where $t_n$ is a shorthand we use throughout this work. Using the Bloch ansatz, we compute the anomalies mentioned in the last section:
\begin{eqnarray}
    \mathcal{A}_{\hat{t}_n} = \mathcal{A}_{\hat{t}_n p} = 0.
\end{eqnarray}
Additionally, $t_n$ is conveniently related to derivatives of itself via $t_n' = \frac{i 2\pi n} {a} t_n$, and we obtain for $n \not = 0$:
\numparts
\begin{eqnarray}
    \langle t_n p \rangle + n \frac{\pi}{a} \langle t_n \rangle = 0, \label{EqTP1}\\
    4n^2 \left( \frac{\pi}{a} \right)^2 \langle t_n p \rangle + 4n \frac{\pi}{a} \langle t_n p^2 \rangle + i \langle t_n V' \rangle = 0,  \\
    \langle t_n p^2 \rangle + \langle t_n V \rangle - E \langle t_n \rangle = 0. \label{EqTP2}
\end{eqnarray}
\endnumparts
For $n = 0$, we have $\langle t_0 \rangle = \langle 1 \rangle$, and the only non-trivial bootstrapping identity at this level is
\begin{eqnarray}
    \langle V'\rangle = 0, \label{Eqn0Identity}
\end{eqnarray}
which is the statement that the average derivative of a periodic potential is zero. For $n \not = 0$; however, the recursion relation must then be derived from the expression
\begin{eqnarray}\label{genrec}
    4n \frac{\pi}{a} \left[ E - n^2 \left( \frac{\pi}{a} \right)^2 \right] \langle t_n \rangle = 4n \frac{\pi}{a} \langle t_n V \rangle - i \langle t_n V' \rangle . \label{EqMaster}
\end{eqnarray}
Equation (\ref{EqMaster}) will mostly be used as the master relation for bootstrapping periodic potentials throughout our work. There are two ways to deal with the derivative of the potential that appears here. As it is periodic, one can complete a Fourier decomposition, and write
\begin{eqnarray}
    V(x) = \sum_{m=-\infty}^{\infty} V_m t_m,
\end{eqnarray}
where the modes are given by
\begin{eqnarray}
    V_m = \frac{1}{a}\int_{-a/2}^{a/2} dx V(x) e^{-i2\pi n x /a} = \frac{1}{a}\int_{-a/2}^{a/2} dx V(x) t_{-m}. \label{EqFourierModes}
\end{eqnarray}
Such a decomposition yields
\begin{eqnarray}
    n \left[ E - n^2 \left( \frac{\pi}{a} \right)^2 \right] \langle t_n \rangle = \sum_{m=-\infty}^{\infty} \left( 2n + m \right) V_m \langle t_{n+m} \rangle. \label{EqRecurrenceFourier}
\end{eqnarray}
Such a formulation of the recursion relation is practical for a potential composed of a finite number of $\lbrace V_m \rbrace$ modes, such as the cosine potential discussed in \cite{tchoumakov_bootstrapping_2021,berenstein_bootstrapping_2022}. In that instance, $V_m \propto \delta_{m,\pm 1}$, and the recurrence relation links each moment to only two others. Indeed, one can reproduce the exact recurrence relation discussed in \cite{berenstein_bootstrapping_2022,tchoumakov_bootstrapping_2021} using our general formula above.
\medskip

However, equation (\ref{EqRecurrenceFourier}) becomes much less useful for potential with a large number of Fourier modes, like in the case for a delta function (where the set of $\lbrace V_m \rbrace$ is infinite). We thus consider a second approach, one that involves reconstructing the probability distribution function from the moments. Starting with (\ref{genrec}) and expanding the expectation value as an integral and integrating by parts we obtain
\begin{eqnarray}
    \langle t_n V' \rangle = - \langle t_n' V \rangle - \int_{-a/2}^{a/2} dx \rho'(x) t_n V,
\end{eqnarray}
where all other contributions have vanished due to the periodicity of $t_n$, $V(x)$, and $\rho(x)$. The recursion relation (\ref{genrec}) then becomes
\begin{eqnarray}
    4n \frac{\pi}{a} \left[ E - n^2 \left( \frac{\pi}{a} \right)^2 \right] \langle t_n \rangle = 2n \frac{\pi}{a} \langle t_n V \rangle + i \int_{-a/2}^{a/2} dx \rho'(x) t_n V. \label{EqRecursionRhoDashed}
\end{eqnarray}
To proceed further, we get a handle on $\rho'(x)$ by using integration by parts to obtain from equation (\ref{Eqn0Identity})
\begin{eqnarray}
    0 = \int_{-a/2}^{a/2} dx \rho(x) V' = - \int_{-a/2}^{a/2} dx \rho'(x) V, \label{EqRhoDashedConstraint}
\end{eqnarray}
which will allow us to compute the recursion relation in the case studied in this work.
\subsection{The recursion relation in the Kronig-Penney Model}
\label{SectionRecursioninKP}
We now turn our attention to the central point of the paper, the Kronig-Penney model \cite{kronig_quantum_1931}, where an infinite array of rectangular barriers approximates a 1D crystalline lattice. In the limit that these barriers become infinitely narrow, they are treated as a series of periodic delta spikes of uniform height, and the potential becomes
\begin{eqnarray}
    V(x) = A \sum_{m=-\infty}^{\infty} \delta \left(x - m a \right), \label{EqKPPotential}
\end{eqnarray}
where $a$ is the spacing between the barriers (the period of the lattice) and $A$ parameterises the height of the potential barriers. Analytic solutions to this system is a textbook problem and solving the Schr\"{o}dinger equation in Fourier space yields the familiar dispersion relation \cite{singh_kronigpenney_1983}
\begin{eqnarray}
    \cos \left( k a \right) = \cos \left( a \sqrt{E} \right) + \frac{A}{2\sqrt{E}} \sin \left( a \sqrt{E} \right), \label{EqKPDispersion}
\end{eqnarray}
where $k$ is the Bloch wavevector introduced earlier. For $V(x)$ as defined in equation (\ref{EqKPPotential}), we can explicitly compute the right hand side of equation (\ref{EqRhoDashedConstraint}) and obtain $\rho'(0) = 0$, which implies that the integral on the right hand side of equation (\ref{EqRecursionRhoDashed}) vanishes; i.e. $\int_{-a/2}^{a/2} dx \rho'(x) t_n V = 0$. We additionally compute that
\begin{eqnarray}
    \langle t_n V \rangle = A \rho(0), 
\end{eqnarray}
so that equation (\ref{EqRecursionRhoDashed}) gives rise to,
\begin{eqnarray}\label{rho0}
    \langle t_n \rangle = \frac{A\rho(0)}{2} \frac{1}{E - n^2 \pi^2/a^2}, \label{EqTN}
\end{eqnarray}
which is valid for $n \not =0 $ and $E \not = n^2 \pi^2/a^2$. In the special case that $E_m = m^2 \pi^2/a^2$ for some $m \in \mathbb{Z}/\lbrace 0 \rbrace$, we can explicitly find that $\rho(0) = 0$, together with $\langle t_n \rangle = 0$, provided we have $n \not = \pm m$ or $n = 0$. This particular limit is exactly the one corresponding to the infinite square well, as will be discussed in detail in \ref{AppInfSquareWell}. Observe that for $E < n^2\pi^2/a^2$, the denominator of (\ref{rho0}) is always negative, and the sign of $\langle t_n \rangle$ is always opposite to that of $\rho(0)$ i.e. if $\rho(0) > 0$, as we would expect, $\langle t_n \rangle$ is negative for all non-zero $n$.
\medskip

Having solved the recursion relation for the moments, we can now reconstruct the wavefunction density $\rho(x)$. As we did with $V(x)$, we write down a Fourier expansion of the probability density:
\begin{eqnarray}
    \rho(x) = \sum_{m=-\infty}^{\infty} \rho_m t_m,
\end{eqnarray}
where the modes are again given by,
\begin{eqnarray}
    \rho_m = \frac{1}{a} \int_{-a/2}^{a/2} dx \rho(x) t_{-m} = \frac{1}{a} \langle t_{-m} \rangle.
\end{eqnarray}
Substituting equation (\ref{EqTN}) for $m \not =0$ and $\langle t_0 \rangle = 1$, we obtain the functional form of the probability density:
\begin{eqnarray}
    \rho(x) = \frac{1}{a} + \frac{1}{a} \sum_{m=1}^{\infty} \frac{A \rho(0)}{E - \pi^2 m^2/a^2} \cos \left( \frac{2\pi m x}{a} \right). \label{EqRHOX}
\end{eqnarray}
From this expression, we can solve for $\rho(0)$. Setting $x = 0$ and using the identity
\begin{eqnarray}
    \cot z = \frac{1}{z} + \sum_{n=1}^{\infty} \frac{2z}{z^2 - n^2 \pi^2}, 
\end{eqnarray}
we obtain
\begin{eqnarray}\label{exprob}
    \rho(0) = \frac{2}{A \left[ \frac{2a}{A} + \frac{1}{E} - \frac{a}{\sqrt{E}} \cot \left( a \sqrt{E} \right) \right]}. \label{EqRHOZERO}
\end{eqnarray}
We plot this solution in Figure\begin{figure}[t!]

\includegraphics[clip,width=12cm]{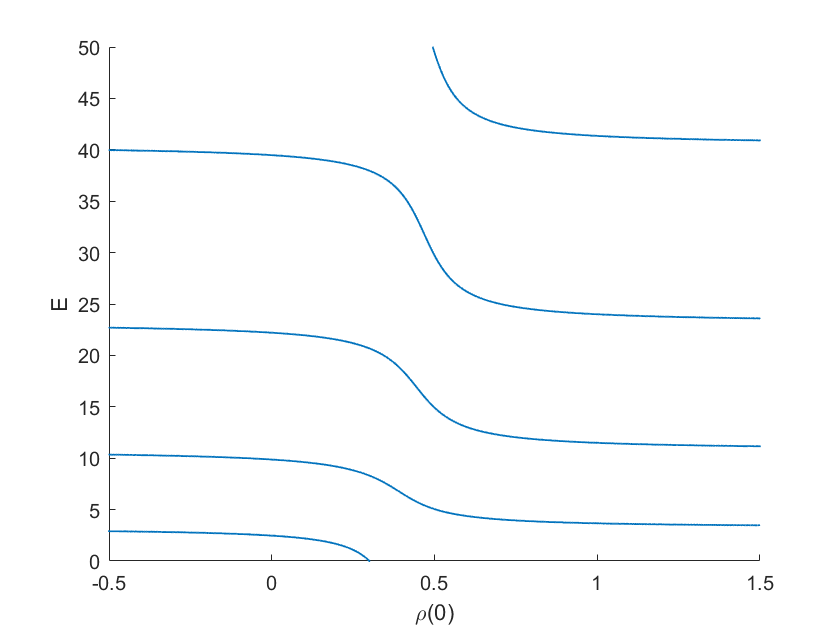}
\centering

\caption{Analytic solution for $E$ as a function of $\rho(0)$ for $a = A = 2$.}
\label{fig:rhozeroanaly}
\end{figure} \ref{fig:rhozeroanaly} for $a = A = 2$. Observe that the function centres on $\rho(0) = 1/a$, and that negative values of $\rho(0)$ are allowed by this expression. This latter fact appears unphysical, and we will return to this subtlety shortly.
\medskip

As noted before, $\rho(0)= 0$ when $E = m^2 \pi^2/a^2$ for $m \in \mathbb{Z}/\lbrace 0 \rbrace$. The poles physically correspond to the limit $a \rightarrow 0$ where it is impossible to define a wavefunction (as the potential is infinite everywhere). Having computed $\rho(0)$ as a function of $E$, equation (\ref{EqTN}) now only depends upon $E$, so numerically, our bootstrapping procedure needs only to consider one parameter.
\section{Numeric implementation of the bootstrap}
\label{SecNumericImplementation}
Having constructed (and in our case, solved) a recursion relation between the moments of the probability distribution, the other key element of the bootstrapping procedure is to restrict the values of those moments via a positivity constraint. In particular, consider a list of linearly independent operators $\mathcal{O}_i$. Then some linear combination $\mathcal{O} = \sum_i a_i O_i$ must obey the constraint \cite{lin_bootstraps_2020,han_bootstrapping_2020}
\begin{eqnarray}
    \left \langle \psi \left| \mathcal{O}^{\dagger} O \right| \psi \right\rangle \geq 0, \label{EqPosConstraint}
\end{eqnarray}
where $\psi$ is an energy eigenstate as before. This constraint of positive definiteness can now be checked numerically, given the initial data from the recursion.
\medskip

As mentioned earlier, previous work suggests using an operator basis with commensurate symmetries to increase the convergence speed of the bootstrap.
%is improved by computing statistical moments that have the same symmetries as the total system. That is, by considering creation and annihilation operators for a diagonalisable system \cite{aikawa_bootstrap_2022}, or periodic moments for a periodic potential \cite{tchoumakov_bootstrapping_2021,aikawa_application_2022,berenstein_bootstrapping_2022}.
Thus, given the operators we have encountered in our periodic potential, we are motivated to consider operators of the form
\begin{eqnarray}
    \hat{O} = \sum_{n=0}^K \sum_{s=0}^L a_{n,s} \hat{p}^s t_n,
\end{eqnarray}
where $a_{n,s}$ is set of $(K+1) \times (L+1)$ arbitrary complex coefficients \cite{tchoumakov_bootstrapping_2021}. In the case, expanding equation (\ref{EqPosConstraint}) yields
\begin{eqnarray}
    \sum_{n,n'=0}^K\sum_{s,s'=0}^L a_{n,s}^* a_{n',s'} \left \langle t_{-n} \hat{p}^{s+s'} t_{n'} \right \rangle \geq 0, 
\end{eqnarray}
which is equivalent to the requirement that the matrix $\mathcal{M}_{n\sigma,m\tau} = \langle t_{-m}p^{\sigma+\tau}t_n \rangle$ be positive semi-definite. If we fix $\sigma + \tau$, we have a $(K+1) \times (K+1)$ matrix. From this matrix, we numerically obtain allowed energies by scanning over the parameter space (in our case, the energy $E$) and ruling out energies which do not correspond to a positive definite matrix. The philosophy of the bootstrapping process is that as $K$ increases, the allowed energies should converge to those obtained from explicitly solving the Schr\"{o}dinger equation. In the case $\sigma+\tau = 0$, the requirement for $\mathcal{M}_{n\sigma,m\tau}$ to be positive semi-definite is equivalent to the solution to the classical trigonometric moment problem (as proven in \cite{caratheodory_uber_1911}) and the matrix is referred to as the Toeplitz matrix.
\medskip

Note that when calculating equation (\ref{EqRHOZERO}), we used the fact that the sum in equation (\ref{EqRHOX}) is over infinite $\langle t_n \rangle$. As $\mathcal{M}_{n\sigma,m\tau}$ is computed for a finite number of modes, in what follows, we numerically compute $\rho(0)$ using the finite sum
\begin{eqnarray}
    \rho(0) = \left( a - A \sum_{m=1}^K \frac{1}{E - \pi^2 m^2/a^2} \right)^{-1},
\end{eqnarray}
which is equivalent to equation (\ref{EqRHOZERO}) in the $K \rightarrow \infty$ limit.
\subsection{Obtaining band gaps from the Toeplitz matrix}
We begin by considering the $\sigma + \tau = 0$ case, fixing $a = A = 2$\footnote{The same patterns would occur for $a \neq A$, we simply study this example for aesthetic reasons.}, and then scan over different values of energy $E$. We plot the result of this process for different values of $K$ in Figure\begin{figure}[t!]

\includegraphics[clip,width=12cm]{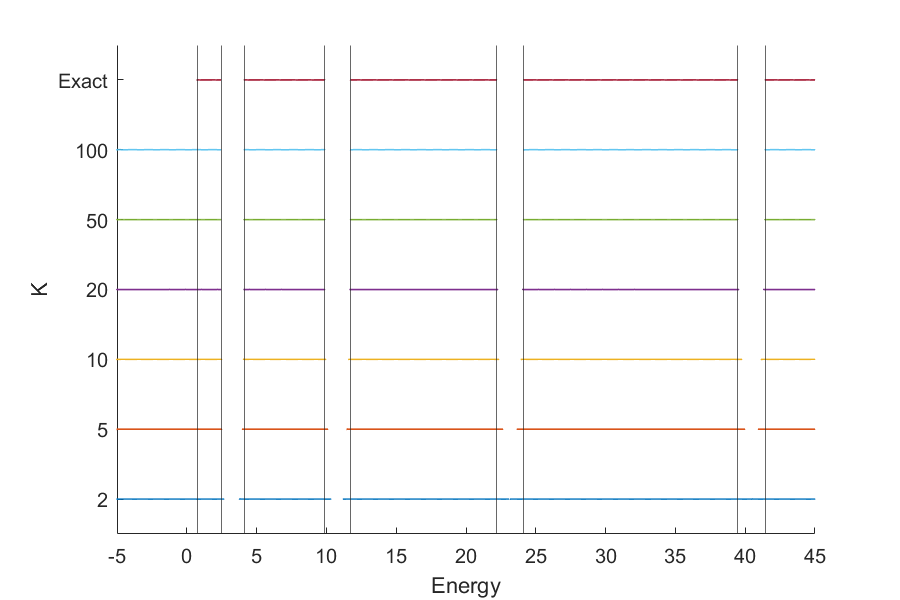}
\centering
\caption{Bootstrap for the Kronig-Penney model at various $K$ and $a = A = 2$. Parallel vertical lines mark the exact energy bands.}
\label{fig:Toeplitz}
\end{figure} \ref{fig:Toeplitz}. The horizontal axis is the energy, and the vertical lines mark the boundaries of the exact energy spectrum from the equation (\ref{EqKPDispersion}). For $K=2$, only the first two band gaps are noticeable, and the numerical band gaps are smaller than the exact value. As $K$ increases, all band gaps become evident and converge to the exact values. We quantify this in Table\begin{table}[t!]
\centering
\begin{tabular}{cc|cccccc|c}
\cline{3-8}
\multicolumn{1}{l}{}                           & \multicolumn{1}{l|}{} & \multicolumn{6}{c|}{K}                                                                                                                           & \multicolumn{1}{l}{}          \\ \cline{3-9} 
                                               &                       & \multicolumn{1}{c|}{2}    & \multicolumn{1}{c|}{5}    & \multicolumn{1}{c|}{10}   & \multicolumn{1}{c|}{20}   & \multicolumn{1}{c|}{50}   & 100  & \multicolumn{1}{c|}{Analytic} \\ \hline
\multicolumn{1}{|c|}{\multirow{3}{*}{$\Delta E$}} & 1st Gap               & \multicolumn{1}{c|}{1.09} & \multicolumn{1}{c|}{1.45} & \multicolumn{1}{c|}{1.58} & \multicolumn{1}{c|}{1.63} & \multicolumn{1}{c|}{1.64} & 1.66 & \multicolumn{1}{c|}{1.66}     \\ \cline{2-9} 
\multicolumn{1}{|c|}{}                         & 2nd Gap               & \multicolumn{1}{c|}{0.90} & \multicolumn{1}{c|}{1.35} & \multicolumn{1}{c|}{1.65} & \multicolumn{1}{c|}{1.80} & \multicolumn{1}{c|}{1.85} & 1.87 & \multicolumn{1}{c|}{1.87}     \\ \cline{2-9} 
\multicolumn{1}{|c|}{}                         & 3rd Gap               & \multicolumn{1}{c|}{0.09} & \multicolumn{1}{c|}{1.03} & \multicolumn{1}{c|}{1.57} & \multicolumn{1}{c|}{1.79} & \multicolumn{1}{c|}{1.91} & 1.93 & \multicolumn{1}{c|}{1.94}     \\ \hline
\end{tabular}
\centering
\caption{First three energy gaps computed numerically at various $K$ and $a = A = 2$. Values are quoted to a numeric error of $\pm 0.01$.}
\label{TabDeltaE}
\end{table} \ref{TabDeltaE}, up to a numeric error of $\pm 0.01$. The band gaps between higher energies converge less rapidly than those between lower energies.
\medskip 

We also plot the values of $\rho(0)$ corresponding to allowed values for $E$ in Figure\begin{figure}[t!]

\includegraphics[clip,width=\columnwidth]{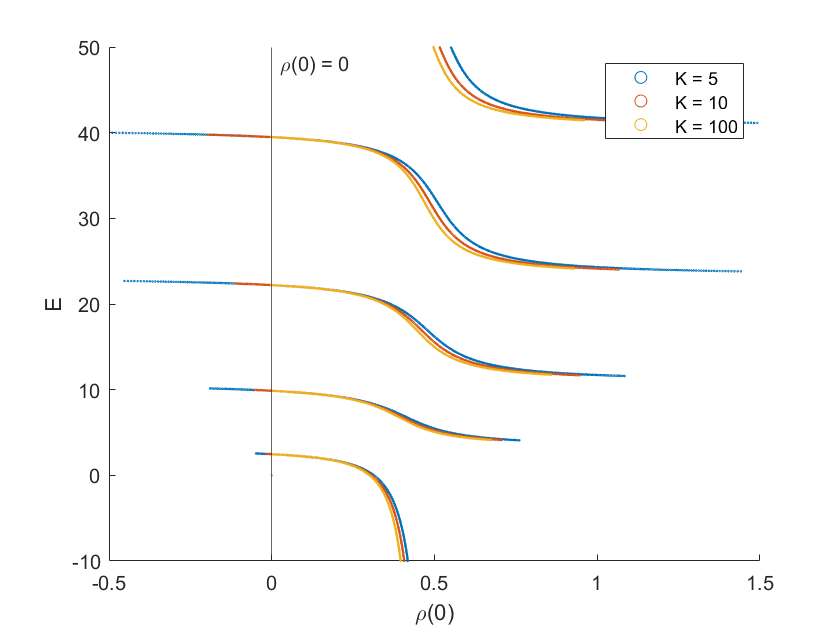}

\caption{Numeric solution for $E$ as a function of $\rho(0)$ for $a = A = 2$ for some values of $K$. Observe that for $K=100$ the negative $\rho(0)$ region stops being allowed, in line with physical intuition.}
\label{fig:EvRZnumer}
\end{figure} \ref{fig:EvRZnumer} for $K =  5$, $10$ and $100$. Given the finite sum, the curves qualitatively differ from the analytic solution shown in Figure \ref{fig:rhozeroanaly} for small $K$. As $K$ increases, the range of allowed $\rho(0)$ decreases; in particular, only positive values of $\rho(0)$ are allowed. This indicates that requiring $\rho(0) \geq 0$ could improve the effectiveness of the bootstrap for small $K$.
\medskip

However, there is an obvious problem: as is evident in Figure \ref{fig:Toeplitz}, requiring the positivity of the Toeplitz matrix does not constrain the minimum energy. Regardless of the value of $K$, all energies below $\pi^2/a^2$ (including negative energies) are allowed by the bootstrap. This indicates that the positivity of the Toeplitz matrix is not a sufficient constraint to find the energy spectrum of the model and that it is necessary to add another tool to our bootstrapping philosophy.
\subsection{Further matrices and constraints for the Kronig-Penney model}
We now move on to the higher constraints, i.e. consider the consequence of enforcing the positivity of the matrix $\mathcal{M}_{n \sigma, m \tau}$ for $\sigma + \tau \geq 1$. We start by constructing elements of the form $\langle t_{-m} p^s t_n \rangle$ from $\langle t_m p^s \rangle$ and the commutation relation 
\begin{eqnarray}
    \left[ p, t_m \right] = \frac{2m \pi}{a} t_m. \label{EqCommut}
\end{eqnarray}
From equations (\ref{EqTP1}) and (\ref{EqTP2}) we immediately have
\numparts
\begin{eqnarray}
    \langle t_n p \rangle = - n \frac{\pi}{a} \langle t_n \rangle, \\
    \langle t_n p^2 \rangle = E \langle t_n \rangle - A \rho(0),
\end{eqnarray}
\endnumparts
and we construct $\langle t_m p^s \rangle$ from equation (\ref{EqFCon2}) by choosing $f = t_m p^s$. In particular, for $s \geq 1$ we obtain (after simplifying with equation (\ref{EqCommut})),
\begin{eqnarray}
    \fl \langle t_n p^{s+2} \rangle = \left( E - 4n^2 \left( \frac{\pi}{a} \right)^2 \right) \langle t_n p^s \rangle - 4n \frac{\pi}{a} \langle t_n p^{s+1} \rangle - \frac{A}{a} \sum_{m=-\infty}^{\infty} \langle t_{m+n} p^s \rangle. \label{EqTNPRecursion}
\end{eqnarray}
A similar double recursion relation was obtained in \cite{tchoumakov_bootstrapping_2021}; however, in that work, the higher order moments (specifically for $n = 0$) were found to be numerically unstable. Analytically computing the next two terms in the recursion, we can easily see the reason:
\begin{eqnarray}
    \langle t_n p^3 \rangle = - n \frac{\pi}{a} \left( E \langle t_n \rangle - 2 A \rho(0) \right), \\
    \langle t_n p^4 \rangle = E^2 \langle t_n \rangle - A \rho(0) \left[ 2E + 4 n^2 (\pi/a)^2 - \frac{A}{a} \sum_{m=-\infty}^{\infty} \right], \label{EqTNP4}
\end{eqnarray}
or to say specifically, the recursion leads to a non-convergent infinite sum.\footnote{The symbol $\sum_{m=-\infty}^{\infty}$ implies an infinite series of $...+1+1+1+...$ for our case.} %\textcolor{red}{Question: For the sake of the reader, should we write down the generic structure of $\langle t_n p^{2+k} \rangle$ here so that the infinite sum is clear to spot?}

We will be discussing a way of analytically treating this sum in Section \ref{SecDispersionRelation}. However, for numerics, here we treat it as a finite size correction by summing over the (finite) number of modes determined by the bootstrap matrix of dimension $K$. With that in mind, in what follows, we consider only the cases where $\sigma + \tau \leq 4$, where the infinite sum first appears, as the first investigation into these finite size effects. To do so, we can analytically construct the following matrices

\begin{eqnarray}
   \fl \langle t_{-m} p t_n \rangle = \frac{\pi(n+m)}{a}\langle t_{n-m} \rangle, \\
    \fl\langle t_{-m} p^2 t_n \rangle = \left( \left( \frac{2\pi}{a} \right)^2 nm + E \right) \langle t_{n-m} \rangle - A \rho(0), \\
    \fl \langle t_{-m} p^3 t_n \rangle = \frac{\pi}{a} \left(n+m\right) \left[ \left[ 3 E  - 2 \left( m^2 - 4mn + n^2 \right) \left( \frac{\pi}{a} \right)^2 \right] \langle t_{n-m} \rangle - 3 A \rho(0) \right], \\
    \fl \langle t_{-m} p^4 t_n \rangle = E^2 \langle t_{n-m} \rangle + E 8n \left( \frac{\pi}{a} \right)^2 \left( n+ 2m \right) \langle t_{n-m} \rangle \\
    - 8n \left[ m^3 -3m^2 n - mn^2 + n^3 \right] (\pi/a)^4 \langle t_{n-m} \rangle \nonumber \\
    - A \rho(0) \left[ 2E + 4(m+n)(m+2n) (\pi/a)^2 - \frac{A}{a} \sum_{m=-\infty}^{\infty} \right]\,, \nonumber
\end{eqnarray}
using the expressions of  $\langle t_n p^{s+2} \rangle$. It is evident that as $\sigma + \tau$ increases, the matrices become more unwieldy. One can check, for larger values of $\sigma + \tau$ we obtain nested infinite sums, and this seemingly divergent correction becomes more unstable and harder to regulate. We will be keeping them as is for now, and in the following section, we will discuss how these unregulated infinite sums contribute to the spectrum in a regulated way. Subsequently, in Section \ref{SecDispersionRelation} we will explicitly regulate these divergences in order to construct a dispersion relation, as we would need to involve the exact probability density (\ref{exprob}) for that case.
\subsection{Constraining the minimum energy}
\begin{figure}[t!]

\subfloat[][$K = 5$]{%
  \includegraphics[clip,width=0.51
 \columnwidth]{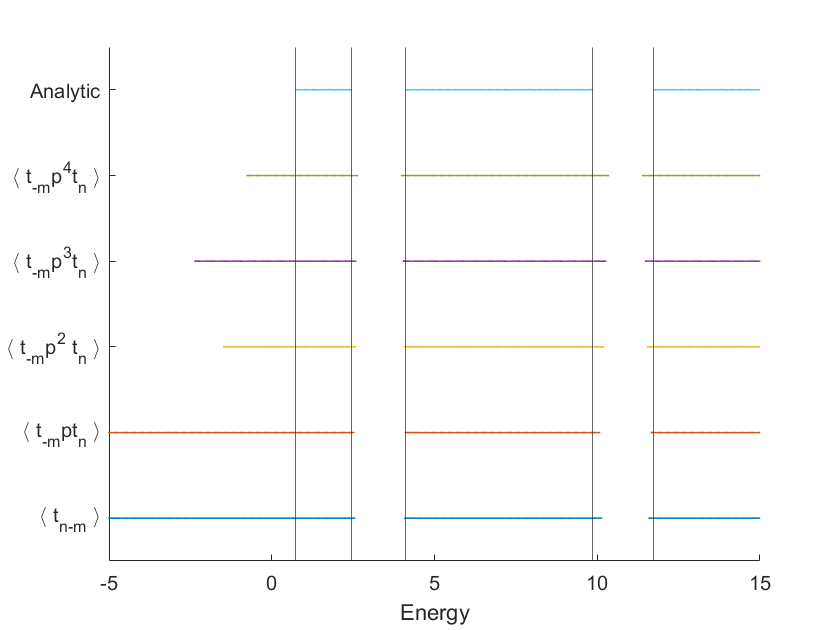} \label{fig:ComparingPA}
}
\subfloat[][$K = 100$]{%
  \includegraphics[clip,width=0.51 \columnwidth]{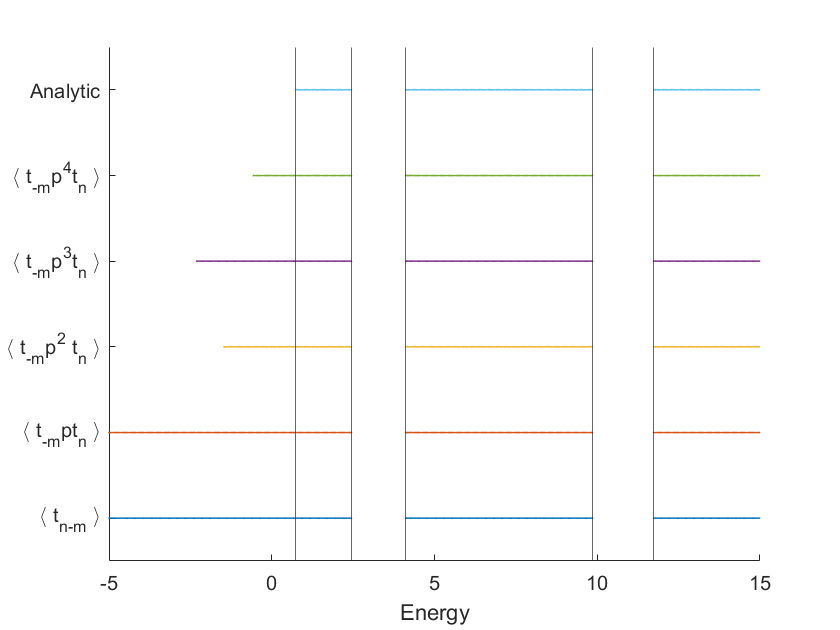} \label{fig:ComparingPb}
}
\caption{Bootstrap for the Kronig-Penney model at $a = A = 2$ for $0 \leq \sigma + \tau \leq 4$ for a) $K = 5$ and b) $K = 100$. Parallel vertical lines mark the exact energy bands. The minimum energy is more constrained for even $\sigma + \tau$ values. }
\label{fig:ComparingPK5K100}
\end{figure}

As discussed earlier, the problem with using the Toeplitz matrix constraint was that it did not constrain the lowest energy value and allowed an infinite sea of negative values. To demonstrate a potential way of improving this using higher constraints, we start with $K = 5$, and fixing $a = A = 2$, we scan over different values of energy $E$ for different $\sigma + \tau$. We plot the result of this process for each of the matrices with $0 \leq \sigma + \tau \leq 4$ in Figure \ref{fig:ComparingPA}.
\medskip

Remember, we are now interested in constraining both the band gaps and the minimum allowed energy. Observe that as $\sigma + \tau$ increases for $K=5$, the band gaps are narrower, i.e. less accurate. This difference becomes negligible as $K$ increases: we plot the result of the bootstrapping process for $K = 100$ in Figure \ref{fig:ComparingPb}, and the exact band gaps are obtained for all $\sigma + \tau$ up to the numerical error of $0.3 \%$. The bootstrapping procedure is most efficient at obtaining the exact band gap structure for smaller $\sigma + \tau$.
\medskip

On the other hand, now consider the behaviour of the minimum energy. For $\sigma + \tau = 1$ a near-continuum spectrum of negative energies is still allowed, but for $\sigma + \tau > 1$, i.e. increasing the power of momentum leads us to a finite minimum bound on the energy. There appears to be an even-odd asymmetry, as evident in Figure \ref{fig:ComparingPA} and \ref{fig:ComparingPb}, where the even powers of momentum in the calculated moment lead to a more accurate constraint than the odd powers. One can interpret this as an artefact of the analytic dispersion relation (equation (\ref{EqKPDispersion})) being only dependent on even powers of Bloch momentum, which is a consequence of the spatial symmetry of the system. We will return to this discussion in Section \ref{SecDispersionRelation}.
\medskip

The best approximation to the minimum energy is achieved for $\sigma + \tau = 4$. This matrix contains a non-convergent infinite sum as it depends on $\langle t_n p^4 \rangle$. In Figure\begin{figure}[t!]
\subfloat[]{
\includegraphics[clip,width=0.52\columnwidth]{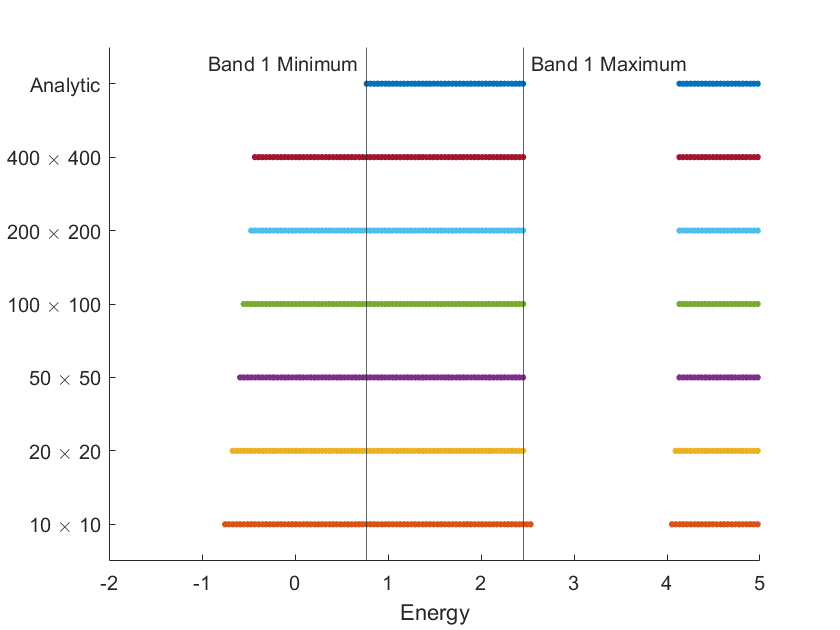}\label{fig:FSPa}}
\subfloat[]{
\includegraphics[clip,width=0.52\columnwidth]{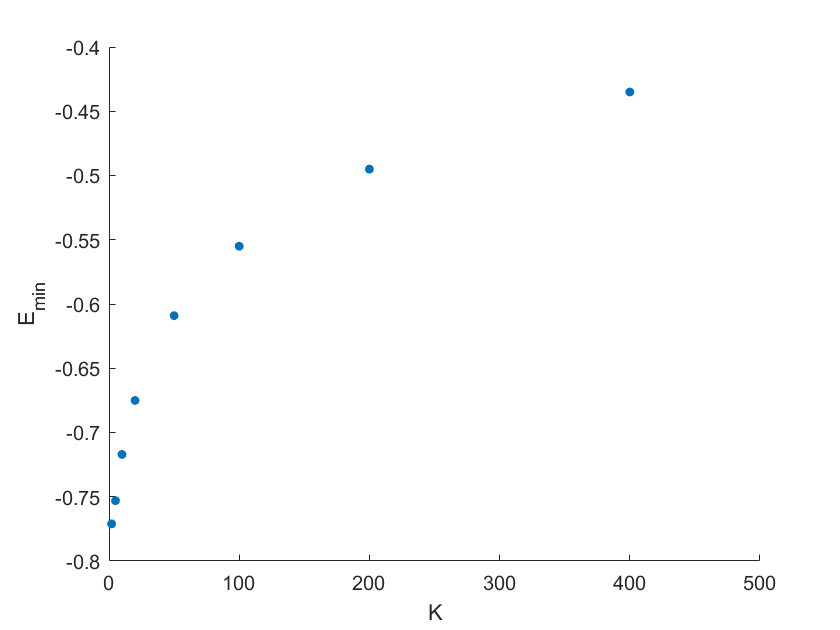}\label{fig:FSCb}}

\caption{ Bootstrap for the Kronig-Penney model at $\sigma + \tau = 4$ and $a = A = 2$ at various $K$, where we zoom into the structure of the first band. The exact minimum energy is marked in a) by a vertical line. In b), we plot the minimum energy obtained numerically as a function of $K$. }
\label{fig:FSC}
\end{figure} \ref{fig:FSPa} we plot the allowed energy levels in the $\sigma + \tau = 4$ case for different values of $K$. We plot the minimum energy as a function of $K$ in Figure \ref{fig:FSCb}. As $K$ increases, the minimum energy increases and converges towards a constant value, which is a lower bound on the minimum energy. This demonstrates that, numerically, the infinite sum contributes a finite size correction, which converges in the limit of a large number of modes.
\medskip

We conclude from Figure \ref{fig:ComparingPK5K100} that the bootstrapping philosophy needs to be updated for periodic potentials. Not only does increasing the value of $K$ increase the accuracy of the energy spectrum obtained, but it is also necessary to explore multiple powers of momentum $p$ (higher values of $\sigma + \tau$). Lower powers of momentum converge more quickly to the desired band gaps, whilst higher powers of momentum converge to provide a better constraint on the minimum energy. This is because, by exploring more powers of momentum, more moments to constrain the dispersion relation are being accessed.
\section{Towards constructing the dispersion relation}
\label{SecDispersionRelation}
As discussed in the introduction, constraining the dispersion relation for periodic systems using the moment recursion relation is much more involved. We sketch here a resolution of the issue using quantities computed in the previous sections. The reader is reminded that a wavefunction of the Bloch form satisfies $\psi(x) = \exp(ikx) u(x)$, applying the exponential of the momentum operator yields $e^{i\hat{p}a}| \psi \rangle = e^{ika} | \psi \rangle$, so the Bloch wavevector $k$ is extracted via $\langle e^{i\hat{p}a} \rangle = e^{ika}$. The dispersion relation is thus extracted from
\begin{eqnarray}
    \cos \left( ka \right) = \sum_{n=0}^{\infty} \left(- 1 \right)^n \frac{a^{2n}}{(2n)!} \langle \hat{p}^{2n} \rangle, \label{EqCOSKADefin}
\end{eqnarray}
where $p^{2n}$ is extracted from the $n = 0$ instance of equation (\ref{EqTNPRecursion}), which reads,
\begin{eqnarray}
    \langle p^{s+2}\rangle = E \langle p^s \rangle- \frac{A}{a} \sum_{m=-\infty}^{\infty} \langle t_m p^s \rangle,
\end{eqnarray}
and $\langle t_m p^s \rangle$ are themselves computed from equation (\ref{EqTNPRecursion}). It was found in \cite{tchoumakov_bootstrapping_2021} that following such a procedure proved numerically unstable; we propose that this is because this leads to larger and larger sums, even for potential with a finite number of Fourier modes $ \lbrace V_m \rbrace$. We instead deal with the infinite sums using two known values of the Riemann zeta function
\begin{eqnarray}\label{EqZeta2n}
    \zeta \left( 0 \right) = -\frac{1}{2}, ~~~~~~~~~~    \zeta \left( - 2n \right) = 0 \qquad& n \in \mathbb{N}. 
\end{eqnarray}
With that, the infinite sum in equation (\ref{EqTNP4}) simplifies, and we find via induction
\begin{eqnarray}
   \fl \langle t_n p^{2s} \rangle = E^s \langle t_n \rangle - A \rho(0) \left( E^{s-1} + \sum_{m=1}^{s-1} c^{(1)}_{m,n} E^{s-1-m} \left( n^2 \pi^2 / a ^2 \right)^m \right), \\
     \fl\langle t_n p^{2s+1} \rangle = - n\frac{\pi}{a} \left[ E^s \langle t_n \rangle - A \rho(0) \left( E^{s-1} + \sum_{m=1}^{s-1} c^{(2)}_{m,n} E^{s-1-m} \left( n^2 \pi^2 / a ^2 \right)^m \right) \right],
\end{eqnarray}

where the $c^{(i)}_{m,n}$ for $i \in \lbrace 1,2 \rbrace$ are numeric factors without an apparent closed form, but those will not prove relevant to our interests. Using equation (\ref{EqZeta2n}), we readily obtain:
\numparts
\begin{eqnarray}
    \langle p^{2s} \rangle = E^s \left( 1 - \frac{s A \rho(0) }{ E} \right), \label{EqP2SRHOZERO} \\
    \langle p^{2s+1} \rangle = E^s \langle p \rangle,
\end{eqnarray}
\endnumparts
where $\langle p \rangle = - a/2 j \left( a/2 \right)$ as computed in terms of the particle current $j$ in \ref{AppComputingExpP} using a proper calculation of the anomaly $\mathcal{A}_x$. We do not solve a recursion relation for the particle current in this work, but by considering different operator forms of $\mathcal{O}$, one could do so. 
\medskip

Substituting the solution from equation (\ref{EqP2SRHOZERO}) back into equation (\ref{EqCOSKADefin}) and using equation (\ref{EqRHOZERO}) for $\rho(0)$ does not help us a lot.
%, we are unable to obtain the exact dispersion relation equation (\ref{EqKPDispersion}).
However, the dispersion relation is a semi-classical quantity. Keeping in with that, in \cite{tchoumakov_bootstrapping_2021} it was noted that the dispersion relation for Bloch momenta could be obtained only by taking the average of a statistical ensemble. In accordance with that, we then focus on the expectation value of $\rho(x)$ instead, which is simply
\begin{eqnarray}
    \langle \rho(x) \rangle = 1/a,
\end{eqnarray}
and then trivially $\langle \rho(0) \rangle = 1/a$. Now, using this expectation value in (\ref{EqP2SRHOZERO}), we get, 
\begin{eqnarray}
    \langle p^{2s} \rangle = E^s\left(1 - s \frac{A}{aE}\right),
\end{eqnarray}
which we can then substitute the above into equation (\ref{EqCOSKADefin}) to obtain
\begin{eqnarray}
    \cos ka = \langle \cos pa \rangle = \cos \left( a \sqrt{E} \right) + \frac{A}{2\sqrt{E}} \sin \left( a \sqrt{E} \right),
\end{eqnarray}
exactly the relation in equation (\ref{EqKPDispersion}). We have thus proposed an analytic approach to deal with the unstable summations present in the recursion for $p^s$ observed in \cite{tchoumakov_bootstrapping_2021} and demonstrated its effectiveness for the Kronig-Penney model. It remains a question for future investigations whether (and how) this trick can be implemented numerically.
\section{Conclusions}
\subsection*{Summary}
In this work, motivated by recent advancements, we explore the effectiveness of the bootstrap method for the Dirac comb model of a 1d lattice system (the Kronig-Penney model). This problem generally can be easily solved analytically, but it is very important to pass it through this algorithm for the sake of further development. We successfully achieved some progress in terms of developing the bootstrap as a tool for solving generic periodic systems using this model as an example. We derived a general dispersion relation for a Bloch periodic problem, considering anomalies generated by the boundary conditions. We also demonstrated the benefit of explicitly reconstructing probability density functions to have analytical handling of the band structure associated with the problem.
\medskip

Numerically, we proposed novel positivity constraints involving a spatially periodic operator basis to zero in on the allowed energy eigenvalues of the system. An intriguing question associated with the Quantum Mechanics bootstrap program is finding constraints which are enough, with a sufficiently sized Hankel matrix, to converge on the exact spectrum of the system. For our case, it turned out that not only a larger size of the constraint matrix but also the power of the momentum used to change the allowed energies pretty drastically, with higher powers generating the band structure better. We augmented our discussion by providing a roadmap toward reconstructing the Kronig-Penney dispersion relation from our analytic considerations.
\subsection*{Comments about the minimum energy}
One intriguing feature of our numerical calculation is the presence of the seemingly unconstrained minimal energy, which we could minimize by using higher powers of momentum. However, there lingers a question about how to constrain this value properly. In a recent work by Morita, \cite{morita_bootstraping_2022}, it was shown that the positive definiteness of the Hamburg matrix for $K = 2$ is equivalent to the Heisenberg uncertainty relation. Computing $\langle x^2 \rangle$ explicitly from equation (\ref{EqRHOX}), requiring $\langle x^2 \rangle \langle p^2 \rangle \geq 1/4$ can be used numerically to obtain finite minimum energy for our model. 
\medskip

A bit of tinkering reveals, for $a = A = 2$, the Heisenberg constraint obtains $E_{min} = 0.245$, as compared to the numerical value $E_{min} = -0.435$ obtained via the $\sigma + \tau = 4$ matrix in the $K = 400$ case, and the analytic value $E_{min} = 0.741$. Whilst the Heisenberg constraint provides a more effective constraint on the minimal value; it allows all energies $E > E_{min}$; that is, it cannot obtain the rest of the energy spectrum. That observation feeds into the narrative, which is apparent from our results; the Toeplitz matrix converges quickly to the band gaps for $E > \pi^2/a^2$, but isn't useful for constraining the minimum possible energy.
\medskip

In this work, we found that the matrices $\mathcal{M}_{n \sigma,m\tau}$ for $\sigma + \tau > 0$ provide one means to get a finite lower bound on the minimum energy (whilst at the same time constraining the band gaps). The Heisenberg relation simply samples another constraint from the set of possible quadratic constraints. It remains an open question of which constraint is most effective for finding the exact minimum energy.
\subsection*{ Comments about a quasi-periodic case} 

In this work, we have considered periodic potentials and focused on the Kronig-Penney model as a particular example. A closely related class of systems are those with quasi-periodic potentials, which have emerged in recent years as important classes of models in condensed matter physics, like the celebrated Aubry-André-Harper (AAH) model \cite{harper_single_1955,aubry_analyticity_1980}. The simplest example of such a quasi-periodic potential in the continuous regime is:
\begin{eqnarray}
    V(x) = \cos \left( \frac{2\pi}{a} x \right) + \cos \left( \alpha \frac{2\pi}{a} x + \beta \right),
\end{eqnarray}
where $\alpha$ is irrational (so that there is no periodic solution) and $\beta$ contributes a phase. We can consider the machinery developed in this work to try and construct a recursion relation for moments for such a potential (although those results were derived for the explicitly periodic case). Computing the Fourier modes $V_m$ to substitute into equation (\ref{EqRecurrenceFourier}), we obtain
\begin{eqnarray}
    \fl V_m = \frac{1}{2} \left[ \delta_{m,1} + \delta_{m,-1} \right]  + \frac{1}{\pi (\alpha^2-m^2)} \left( -1 \right)^m \sin \left( \pi \alpha \right) \left[ m \cos \beta - i \alpha \sin \beta \right],
\end{eqnarray}
so we can see the set $\lbrace V_m \rbrace$ is infinite in size. This would require an infinite sum in the recursion relation, meaning that it is not numerically solvable. Alternatively, one may explicitly compute
\begin{eqnarray}
    \langle t_n V \rangle = \frac{1}{2} \left \langle t_{1-n} + t_{-1-n} + e^{-i\beta} e^{i \frac{2\pi}{a} \left( \alpha - n \right)} + e^{i\beta} e^{i \frac{2\pi}{a} \left( - \alpha - n \right)} \right \rangle.
\end{eqnarray}
Observe that, as $\alpha$ is irrational, the terms dependent on it are outside the sequence of $t_n$, and it is not possible to rescale the operator basis to include them in the sequence. Furthermore, to conclude the vanishing of anomalies, we assumed an exactly periodic (Bloch) wavefunction density, which is no longer necessarily valid. A probable way out is to consider an explicitly quasi-periodic basis of operators and rework the whole algorithm. This is beyond the scope of this work and will be discussed in future communication. 
%We are currently investigating ways to address these issues.
\subsection*{Future directions}
The bootstrap technique has already proven to be of much interest and seemingly very powerful in solving Quantum Mechanical problems. Periodic problems, especially ones with Bloch symmetry, are very new to this growing list of explorations. We considered one of the simpler examples in this work, and extending our results to other tight-binding models makes perfect sense.
\medskip

Of course, other tractable Quantum Mechanical problems are open to exploration using similar methods. Recently \cite{cho_bootstrapping_2022} has initiated an intriguing study of infinite lattice Ising models in different dimensions using similar bootstrap techniques. One could think of these as a viable alternative to the conformal bootstrap procedure, and this leaves ample scope for improvements in the formalism employed, given the success of the conformal bootstrap program. We hope to contribute to these refinements in future work.

\section*{Acknowledgements}
The authors would like to thank Jyotirmoy Bhattacharya and Diptarka Das for their comments on the draft. MJB was supported by an internship at the Quantum Gravity Unit of Okinawa Institute of Science and Technology Graduate University (OIST) during the course of this project. MJB would like to thank Yasha Neiman for discussions. AB is supported by Mathematical Research Impact Centric Support Grant (MTR/2021/000490), Start-Up Research Grant (SRG/2020/001380) by the Department of Science and Technology Science and Engineering Research Board (India) and Relevant Research Project grant (202011BRE03RP06633-BRNS) by the Board Of Research In Nuclear Sciences (BRNS), Department of atomic Energy, India. The work of ArB is supported by the Quantum Gravity Unit of the OIST. ArB would like to thank TU Wien and Ecole Polytechnique, Paris for kind hospitality during the course of this project. 

\appendix

\section{Bootstrapping the Infinite Square Well}
\label{AppInfSquareWell}
The infinite square well is a standard problem in undergraduate quantum mechanics, considering a potential of the form
\[
  V(x) = \left\{\begin{array}{@{}l@{\quad}l}
      0 & \mbox{if $0 \leq x \leq a$} \\[\jot]
      \infty & \mbox{elsewhere}
    \end{array}\right.
\]
Solving the Schr\"{o}dinger equation, one can find the position-space energy eigenfunctions easily. However, here we will use our moment recursion method to solve the system.
%\begin{eqnarray}
%    \psi_n(x) = \sqrt{\frac{2}{a}} \sin \left( \frac{n\pi x}{a} \right), \label{EqAppISWSol}
%\end{eqnarray}
%with energy $E_n = n^2 \pi^2/a^2$.

To bootstrap the infinite square well, we will consider a basis of operators with $\mathcal{O} = x^n$ and $\mathcal{O} = x^n p$. Over a finite interval $x \in \lbrace 0, a \rbrace$, where the particle is free, these operators generate non-zero anomaly terms. Assuming the position-space wavefunction $\psi(x)$ to be real, we compute the anomalies
%\numparts
\begin{eqnarray}\label{anoISW}
    \fl \mathcal{A}_{x^n} &= n a^{n-1} \psi(a)^2 - \delta_{n,1} \psi(0)^2, \\
    \fl \mathcal{A}_{p} &= i \left[ \psi(a) \left( E - V(a) \right) - \psi(0) \left( E - V(0) \right) + \psi'(a)^2 - \psi'(0)^2 \right], \\
    \fl \mathcal{A}_{x^np} &= i \delta_{n,1} \psi(0) \psi'(0) - i \left[ \psi(a) n a^{n-1}\psi'(a) - \psi(a) a^n \left( E - V(a) \right) - a^n \psi'(a)^2 \right].
\end{eqnarray}
%\endnumparts

In the infinite square well, we assume the Dirichlet boundary conditions $\psi(0) = \psi(a) = 0$, so $\mathcal{A}_{x^n}$ as defined above identically vanishes and we are left with
%\numparts
\begin{eqnarray}
    \mathcal{A}_p &= i a^n \psi'(a)^2, \\
    \mathcal{A}_{x^n p} &= i \left( \psi'(a)^2 - \psi'(0) \right)^2, \qquad& n \not = 0.
\end{eqnarray}
%\endnumparts 
Accounting for these anomalies and noting that $\langle V(x) \rangle = 0$ for the infinite square well, from general recurrence relations (\ref{EqFCon1}), (\ref{EqFPCon1}), and (\ref{EqFCon2}) we obtain for $n \not = 0$:
%\numparts
\begin{eqnarray}
   & \langle X^{n-1} P \rangle = \frac{i}{2} \left( n-1 \right) \langle X^{n-2} \rangle, \\
    &\langle X^n P^2 \rangle = E \langle X^n \rangle, \\
    &- n (n-1) \langle X^{n-2} P \rangle - 2 i n \langle X^{n-1} P^2 \rangle + i a^n \psi'(a)^2 = 0.
\end{eqnarray}
%\endnumparts
Equation (\ref{EqFPCon1}) for $n = 0$ implies that
\begin{eqnarray}
    \mathcal{A}_{x^n p} = 0 \Rightarrow \psi'(a)^2 = \psi'(0)^2,
\end{eqnarray}
and our effective recursion relation for the system becomes
\begin{eqnarray}
    \langle X^n \rangle = - \frac{n(n-1)}{4E} \langle X^{n-2} \rangle + \frac{a^{n+1}}{2 E(n+1)} \psi'(a)^2.
\end{eqnarray}
Fixing $\langle X^0 \rangle = 1$, we obtain
\begin{eqnarray}
    \psi'(a)^2 = \frac{2E}{a},
\end{eqnarray}
%(which agrees with equation (\ref{EqAppISWSol})) 
and the recursion relation then becomes
\begin{eqnarray}
    \langle X^n \rangle = - \frac{n(n-1)}{4E} \langle X^{n-2} \rangle + \frac{a^{n}}{(n+1)}.
\end{eqnarray}
Observe that $\langle X \rangle = a/2$ which is expected from the usual symmetry intuition of the infinite well. 
%, which is again in agreement with equation (\ref{EqAppISWSol}). 
The recursion relation can now be solved by 
%\numparts
\begin{eqnarray}
    \langle X^{2m} \rangle &= \left( 2m \right)! \left( - \frac{1}{4E} \right)^m \sum_{k=0}^m \frac{\left( - 4 E a^2 \right)^k}{(2k+1)!}, \label{AppISWX2m} \\
    \langle X^{2m+1} \rangle &= \left( 2m + 1 \right)! \left( - \frac{1}{4E}\right)^m \sum_{k=0}^{m} \frac{\left( -4 E \right)^k}{(2k+2)!} a^{2k+1}. \label{AppISWX2mP1}
\end{eqnarray}
%\endnumparts
Following a similar procedure to Section \ref{SectionRecursioninKP}, for a function defined over an interval $\left[ 0, a \right]$, we can write a fourier decomposition: 
\begin{eqnarray}
    \rho(x) = \sum_{n=-\infty}^{\infty} c_n \exp \left( \frac{i 2\pi n x}{a} \right),
\end{eqnarray}
and calculate the modes as:
\begin{eqnarray}
    c_n = \frac{1}{a} \int_0^a dx \rho(x) \exp \left( - \frac{i 2\pi n x}{a} \right)
\end{eqnarray}
From equations (\ref{AppISWX2m}) and (\ref{AppISWX2mP1}), we can see that only for the known values of infinite well energies, i.e. for $E = m^2 \pi^2/ a^2$ for $m \in \mathbb{Z}$ we can get:
\begin{eqnarray}
    c_0 = 1/a, \\
    c_{\pm m} = - 1/2a, \\
    c_{n} = 0 \qquad& n \not = \pm m, 0.
\end{eqnarray}
We thus obtain, for $E = m^2 \pi^2/a^2$, and the probability density is:
\begin{eqnarray}\label{EqAppISWSol}
    \rho(x) = \frac{2}{a} \sin^2 \left( \frac{n\pi x}{a} \right),
\end{eqnarray}
in agreement with the solution we directly compute from the Schrödinger wavefunction. One can clearly see that for $E \not = m^2 \pi^2/a^2$, $c_n = 0$, i.e. these energies are not allowed by the constraints (as expected). Note that this consistent solution would never have been possible without the use of proper anomalies in the finite domain. 
\subsection*{Agreement with $\rho(0)$ limit}
We note in Section \ref{SectionRecursioninKP} that, for the Kronig-Penney mdoel, when $E = n^2 \pi^2/a^2$ we have $\rho(0) = 0$. In that case, the sum on the right hand size of equation (\ref{EqRHOX}) has a $0/0$ term for $m = n$. Recall that $\rho(0)$ is itself a function of $E$; and we can compute via L'Hospitals rule that
\begin{eqnarray}
    \lim_{E \rightarrow n^2 \pi^2/a^2} \frac{A\rho(0)}{E - n^2 \pi^2 / a^2} = \frac{\frac{\partial}{\partial E} \left[ A \rho(0) \right]_{E = n^2 \pi^2/a^2}}{\frac{\partial}{\partial E} \left[ E - n^2 \pi^2/a^2 \right]_{E = n^2 \pi^2/a^2}} = - 1,
\end{eqnarray}
and thus in this limit
\begin{eqnarray}
    \rho(x) = \frac{1}{a} \left( 1 - \cos \left( \frac{2\pi nx}{a} \right) \right) = \frac{2}{a} \sin^2 \left( \frac{n\pi x}{a} \right),
\end{eqnarray}
again in agreement with equation (\ref{EqAppISWSol}). This is equivalent to showing that the $\langle t_n \rangle$ for the Kronig-Penney model agrees with the $\langle t_n \rangle$ for the infinite square well in the $E \rightarrow n^2 \pi^2/ a^2$ limit.
\section{Computing $\langle p \rangle$}
\label{AppComputingExpP}
From equation (\ref{EqFCon1}), considering $f = \hat{x}$ we have that
\begin{eqnarray}
    \langle p \rangle = \frac{1}{2i} \mathcal{A}_x.
\end{eqnarray}
It is straightforward to compute, using equation (\ref{EqAnomalyDefinition}) and $\psi (x) = \psi (x+a)$, that
\begin{eqnarray}
    \mathcal{A}_x = a \left[ \psi^*(x) \frac{\partial \psi(x)}{\partial x} - \frac{\partial \psi^*(x)}{\partial x} \psi(x) \right] \left( \frac{a}{2} \right) = ia j \left( \frac{a}{2} \right),
\end{eqnarray}
where $j(x)$ is the particle current defined $j = -i \left( \psi^*(x) \partial_x \left( \psi (x) \right) - \partial_x \left( \psi^*(x) \right) \psi(x) \right)$. We thus have
\begin{eqnarray}
    \langle p \rangle = \frac{a}{2} j \left( \frac{a}{2} \right),
\end{eqnarray}
as quoted in Section \ref{SecDispersionRelation}. Note that  \cite{tchoumakov_bootstrapping_2021} claim $\langle p \rangle = 0$; this is because in that work the authors did not account for anomalies. 
\section*{References}
\bibliographystyle{utphysmodb}
\bibliography{Bootstrapping.bib}

\end{document}